\documentclass[prl,twocolumn,preprintnumbers,nofootinbib,showpacs,amsmath,amssymb,superscriptaddress]{revtex4-1}
\usepackage{psfrag,graphicx}

\usepackage[
      colorlinks=true,
      linkcolor=blue,
      urlcolor=blue,
      filecolor=black,
      citecolor=blue,
            ]{hyperref}
\usepackage{color}
\usepackage{amsfonts}
\usepackage{amssymb}
\usepackage{epstopdf}
\usepackage{dsfont}
\usepackage{epsfig}

\begin{document}
\title{Paramagnetism-Ferromagnetism Phase Transition in a Dyonic Black Hole}
\author{Rong-Gen Cai}
\email{cairg@itp.ac.cn}
\affiliation{State Key Laboratory of Theoretical Physics,Institute of Theoretical Physics,\\
 Chinese Academy of Sciences,Beijing 100190, China \\
King Abdulaziz University, Jeddah, Saudi Arabia}

\author{Run-Qiu Yang}
\email{aqiu@itp.ac.cn}
\affiliation{State Key Laboratory of Theoretical Physics,Institute of Theoretical Physics,\\
 Chinese Academy of Sciences,Beijing 100190, China.}

\begin{abstract}
 Coupling an antisymmetric tensor field to the electromagnetic field in a dyonic Reissner-Nordstr\"om-AdS black hole background, we build a holographic model for the paramagnetism/ferromagnetism phase transition. In the case of zero magnetic field, the time reversal symmetry is broken spontaneously and spontaneous magnetization happens in low temperatures. The critical exponents are in agreement with the ones from mean field theory. In the case of nonzero magnetic field, the model realizes the hysteresis loop of single magnetic domain and the magnetic susceptibility satisfies the Curie-Weiss law.
\end{abstract}
\maketitle

\noindent

{\emph {Introduction.}}--The gauge/gravity duality relates a weak coupling gravitational theory in  a $(d+1)$-dimensional asymptotically anti de Sitter (AdS) space-time to a $d$-dimensional strong coupling conformal field theory (CFT) in the AdS boundary~\cite{Gubser:1998bc,Maldacena:1997re,Witten:1998qj,Witten:1998qj2}. In recent years, many
fascinating condensed matter phenomena have been found using the duality, including holographic superfluids (superconductors)~\cite{Gubser,Hartnoll:2008vx}, (non-)Fermi liquids~\cite{Lee:2008xf,Liu:2009dm,Cubrovic:2009ye}, Josephson junctions~\cite{Horowitz:2011dz,Kiritsis:2011zq,Wang}, superconducting quantum interference device~\cite{Cai:2013sua},  magnetic properties in superconductor~\cite{Montull:2009,Donos:2012yu,Albash:2008eh,Montull:2012fy}, and quantum phase transitions~\cite{Iqbal:2010eh}.

Magnetism is ubiquitous in many strongly correlated electronic systems such as high temperature superconductors and heavy fermion metals, and plays a central role in quantum phase transitions in these materials. The gauge/gravity duality provides new approaches and perspective to understanding these challenging systems.
Yet in holographic contexts, due to various technical challenges, models of magnetism are
scarce and not extensively explored (see e.g.~\cite{Iqbal:2010eh}). In this paper we propose a simple holographic model for magnetism, and present an example of the paramagnetism/ferromagnetism phase transition. We believe the model proposed here should serve as a good starting to explore more complicated magnetic phenomena and
quantum phase transitions.

{\emph {The model.}}--The paramagnetism/ferromagnetism phase transition breaks the time reversal symmetry spontaneously in low temperatures (if spatial dimension is more than 2, it also breaks spatial rotation symmetry). So we need a real field to describe this symmetry breaking spontaneously. Note that in the mean field theory of paramagnetism/ferromagnetism phase transition, the magnetization as the order parameter is a pseudovector and is coupled to the external magnetic field; and in the weak magnetic field, the magnetization is proportional to the magnetic field in magnetic materials. Therefore in the spirit of AdS/CFT correspondence it is natural to introduce a real antisymmetric tensor in bulk, its space-space component describes the dual field of magnetization.  Thus we consider a simple model with action as
\begin{equation}\label{LM2}
S=\frac1{2\kappa^2}\int d^4x\sqrt{-g}(R+\frac6{L^2}-F^{\mu\nu}F_{\mu\nu}-4j^\mu A_\mu+\lambda^2 L_M)£¬
\end{equation}
where
\begin{equation}\label{LM3}
\begin{split}
L_M&=-\frac14\nabla^\mu M^{\nu\tau}\nabla_\mu M_{\nu\tau}-\frac{m^2}4M^{\mu\nu}M_{\mu\nu}\\
&-\frac12M^{\mu\nu}F_{\mu\nu}-\frac J8 V(M_{\mu\nu}).
\end{split}
\end{equation}
Here $2\kappa^2=16\pi G$ and $G$ is the Newtonian gravitational constant, $\lambda$ and $J$~are two constants, $m$ is the mass of the real tensor field $M_{\mu\nu}$, $A_\mu$ is the gauge potential of $U(1)$ gauge field and $j^\mu$ is the U(1) current coming from some field carrying U(1) charge. We here introduce $j^\mu$ into the action for the convenience of discussion. In fact we will set $j^\mu=0$ in this paper. The antisymmetric tensor field $M_{\mu\nu}$~is the effective polarization tensor of the U(1) gauge field strength~$F_{\mu\nu}$, whose physical meaning will be explained shortly. $V(M_{\mu\nu})$ describes the self-interaction of the polarization tensor, which should be expanded as the even power of $M_{\mu\nu}$. The lowest order has two linear independent forms. We take $V(M_{\mu\nu})={M_\mu}^\nu {M_\nu}^\tau {M_\tau}^\delta {M_\delta}^\mu$ for simplicity. The probe limit corresponds to $\lambda\rightarrow0$. The equations of motion for matter fields are
\begin{equation}\label{eomFM}
\begin{split}
   \nabla^\mu F_{\mu\nu}=\frac{\lambda^2}{4}\nabla^\mu M_{\mu\nu}+j^\mu, \\
  \nabla^2M_{\mu\nu}-m^2M_{\mu\nu}-J{M_\mu}^\delta {M_\delta}^\tau {M_\tau}_\nu-F_{\mu\nu}=0.
\end{split}
\end{equation}
From the equations for gauge field, one can see that the polarization field can contribute U(1) current and charge density. If introduces an auxiliary field such as~$H_{\mu\nu}=F_{\mu\nu}-\lambda^2M_{\mu\nu}/4$,  the U(1) current is only associated with the field $H_{\mu\nu}$. This is very similar to what happens for electrodynamics in medium, where the auxiliary fields {\bf D} and {\bf H} are introduced  corresponding to {\bf E} and {\bf B}, respectively.

From equations~\eqref{eomFM} one can see that $j^\mu$ is still a conserved current and associated conserved charge is just the U(1) charge. In the planar symmetry case, its density can be expressed as
\begin{equation}\label{freeQ}
Q=\int_{r\rightarrow\infty}{^*(F_{\mu\nu}-\lambda^2M_{\mu\nu}/4)}=\lim_{r\rightarrow\infty}r^2(F_{tr}-\lambda^2M_{tr}/4).
\end{equation}

In this paper, we will work in the probe limit and the background is a planar dyonic Reissner-Nordstr\"om-AdS black hole. The background geometry and gauge potential take the following forms~\cite{Cai}
\begin{eqnarray}\label{geom}
  ds^2 &=& r^2(-f(r)dt^2+dx^2+dy^2)+\frac{dr^2}{r^2f(r)},\nonumber \\
   f(r) &=&1-\frac{1+\mu^2+B^2}{r^3}+\frac{\mu^2+B^2}{r^4},\\
   A_\mu &=& \mu(1-1/r)dt+Bxdy, \nonumber
\end{eqnarray}
where $\mu$ is the chemical potential, $B$ is the magnetic field, and the AdS radius $L$ has been set to be unity and the black hole horizon $r_+=1$. Then the temperature of the black hole is
\begin{equation}\label{tem}
T=(3-\mu^2-B^2)/4\pi.
\end{equation}
 This bulk magnetic field can be regarded as the external magnetic field in the dual boundary theory. In this paper we will work in grand canonical ensemble, where the chemical potential will be fixed as a nonzero value, while the magnetic field can be taken zero or not. We consider a self-consistent ansatz for the polarization field as
\begin{equation}\label{Mcomp}
M_{\mu\nu}=-p(r)dt\wedge dr+\rho(r)dx\wedge dy.
\end{equation}
Then the non-trivial equations for the polarization field in the background (\ref{geom}) read
\begin{equation}\label{rhop}
\begin{split}
\rho''+\frac{f'\rho'}f-\left( \frac{2f'}f+\frac4{r^2}+\frac{m^2}{r^2f}\right)\rho+\frac{J\rho^3}{r^6f}-\frac{B}{r^2f}=0,\\
p''+\left(\frac{f'}f+\frac4r\right)p'-\left(\frac2{r^2}+\frac{m^2}{r^2f}\right)p-\frac{Jp^3}{r^2f}-\frac{\mu}{r^4f}=0,
\end{split}
\end{equation}
 where a prime denotes the derivative with respect to $r$.  It is interesting to see that these two equations decouple from each other in this case. At the horizon, we impose the regularity condition, which implies the following relation
\begin{equation}\label{expand1}
\begin{split}
&\rho'=2\rho-\frac{\rho(J\rho^2-m^2)-B}{4\pi T},\\
&p'=\frac{Jp^3+m^2p-\mu}{4\pi T}.
\end{split}
\end{equation}
Thus once given the initial values of $\rho$ and $p$ at the horizon, one can integrate equations~\eqref{rhop} to get the solution.

{\emph {The asymptotic behavior and free energy.}}--
In order to keep the background to be asymptotic AdS, one has to require
suitable asymptotic behaviors for $\rho$ and $p$ near the AdS boundary. 
There are two different situations in terms of the value of $m^2$. When $m^2=-4$, a logarithmic term appears in the asymptotic expansion of the solution as $r \to \infty$. We will not consider this case in this paper. When $m^2\neq-4$, we have the following asymptotic behaviors
\begin{equation}\label{asym2}
\begin{split}
\rho\sim \rho_+r^{(1+\delta)/2}+\rho_-r^{(1-\delta)/2}-\frac{B}{4+m^2},\\
p\sim p_+r^{(-3+\delta)/2}+ p_-r^{(-3-\delta)/2}-\frac{\mu}{(4+m^2)r^2},
\end{split}
\end{equation}
where $\rho_{\pm}$ and $p_{\pm}$ are all constants and $\delta=\sqrt{17+4m^2}$. The Breitenlohner-Freedman (BF) bound requires $m^2>-17/4$~\cite{BF1,BF2}. Note that there is an other restriction on $m^2$ coming from the U(1) charge density in~\eqref{freeQ}. A well defined probe limit demands that the charge density contributed by the polarization should be finite and the leading order is $\lambda^2$ order. In order to check this, we have to treat $\lambda^2$ as a samll quantity and solve the full equations of motion of~\eqref{LM2} order by order. However, since the charge density~\eqref{freeQ} is calculated at $r\rightarrow\infty$, the asymptotic AdS boundary guarantees that we need not to solve metric if we only care for the order $\lambda^2$.  A direct calculation gives the contribution coming from the polarization in~\eqref{freeQ} is
\begin{equation}\label{curr2}
Q_M=\frac{\mu\lambda^2}{4(4+m^2)}-\frac{\lambda^2}{4} \left[p_+r^{(1+\delta)/2}+ p_-r^{(1-\delta)/2}\right]+O(\lambda^4).
\end{equation}
 We see that if $p_+\neq0$ or $\delta<1$, the quantity is divergent, which leads the system to be inconsistent. In order to avoid this, we further impose
\begin{equation}\label{m1}
p_+=0, ~~m^2>-4.
\end{equation}
On the other hand, in order to the condensate happens spontaneously when the temperature is lowered, the $m^2$ of the real tensor field
should violate the BF bound of the field in $AdS_2\times R^2$. This leads to the constraint of $m^2$ as
\begin{equation}\label{m2}
-4<m^2<-\frac32.
\end{equation}
Note that here $AdS_2 \times R^2$ is the near horizon geometry for an extremal black hole (\ref{geom}) with a vanishing temperature. Therefore the real tensor field $M$ does not violate its BF bound in $AdS_4$, but does violate in $AdS_2\times R^2$. As a result, the spontaneous condensate is an effect of temperature, it is not sensitive to the details of the self-interaction terms for the tensor field $M$.

The free energy of the dual boundary theory can be obtained by calculating the on shell action of bulk in the Euclidean sector. However,
for a generality and a comparison with the Ginzburg-Landau theory, we here present the off-shell free energy of the dual field theory. In other words, we calculate the bulk action when the equations of motion are no longer required to hold. In the probe limit, a direct calculation gives the free energy density of the polarization tensor as
\begin{equation}\label{Lprho}
\begin{split}
 {\cal G}_{\rm offshell}/\lambda^2  & =
\left.\left(\frac 12f\rho'\rho-\frac{f\rho^2}{r}\right)\right|^{r\to \infty}_{r=1} \\
&~~ + \int_1^\infty dr\left ( \frac{1}{2}\rho \hat{L} \rho +\frac{B\rho}{r^2}-\frac{J\rho^4}{4r^6} \right ),
\end{split}
\end{equation}
where we have set $2\kappa^2=1$ and the operator $\hat L$ has been
defined as $\hat{L}= -\frac{d}{dr} \left (f(r) \frac{d}{dr}\right) + 2\frac{f'}{r}+\frac{4 f+m^2}{r^2}$.
Note that here the contribution coming from the component $p$  to the free energy did not included because it is not relevant to the following discussions. In the on shell case with
the equation of motion, $\hat{L} \rho= J\rho^3/r^6 -B/r^2$,  ${\cal G}_{\rm onshell} $ diverges if $\rho_+\neq0$.  This divergency coming from the boundary terms can be canceled by adding some suitable surface counter term, while the one from the volume integration part cannot be renormalized in the case of $B\neq0$. So in order to make the on shell action finite, we need
\begin{equation}\label{delta1}
\rho_+=0.
\end{equation}
In this case, the boundary terms appearing in the action vanish.
The condition (\ref{delta1}) can also be understood as follows. In the AdS/CFT correspondence, when $B=0$, $\rho_+$ and $\rho_-$ correspond
to the source and vacuum expectation value of dual operator in the boundary field theory (up to a normalization factor), respectively. Therefore one should take $\rho_+=0$ since one wants the condensation to happen spontaneously.

In the on shell case, the free energy density for the polarization field can be written as
\begin{equation}\label{Gprho1}
{\cal G}_{\rm onshell}=\frac {\lambda^2}2\int_1^\infty dr\left (\frac{J\rho^4}{2r^6}\right )-BN,
\end{equation}
where
\begin{equation}\label{QN1}
N=-\frac {\lambda^2}2\int_1^\infty dr\frac{\rho}{r^2}.
\end{equation}
According to this free energy density, the term proportional to $J$ can be interpreted as the self-interacting energy of polarization, and $N$ can be explained as the magnetic moment of the dual boundary field theory.

The spontaneous magnetization in low temperature corresponds to the spontaneous condensate of $\rho$ in the bulk in the absence of external magnetic field. One can see from ~\eqref{Gprho1} that once the spontaneous magnetization phase appears it is always more favored than the trivial solution without the condensation  if $J$ is negative. On the other hand, from the first equation of (\ref{rhop}), the term proportional to $J$ likes the $-\lambda \phi^4$ term in the Higgs mechanism, one therefore requires $J<0$ in order the symmetry to be broken spontaneously.

%
%

To calculate the off shell free energy, let us consider the following Sturm-Liouville eigenvalue problem
\begin{equation}
\hat {L} \rho_n (r)  = \lambda_n w(r)\rho_n(r),
\end{equation}
with boundary conditions:  $\rho_n(r_+)$ is finite at horizon and $\rho_n(\infty)=0$ at $r=\infty$.
Here $\lambda_n$'s are eigenvalues and the weight function $w(r)$ can be an arbitrary positive continuous function in the region $r \in [r_+, \infty)$ so that
the eigenvalue $\lambda_n$ will not influence the asymptotical form of the solution $\rho_n$ at the infinity and $\rho_n$ is square integrable in the region $r \in [r_+, \infty)$. For example, we may take
$w(r)=1/r^k$ with $k>2$  for simplicity in our case. Thus any function, $h(r)$, which is finite and square integrable in the $r \in [r_+, \infty)$, can be expanded as
$h(r) = \sum_{n=1}^{\infty} c_n\rho_n(r),$
where $c_n= \langle \rho_n,h\rangle = \int^{\infty}_{r_+}w(r) \rho_n(r) h(r)dr$. Applying this method to our case, we can show that near the critical temperature $T_c$, the free energy density can be expressed as
\begin{equation}
\label{Gprho6}
{\cal G}_{\rm offshell}/\lambda^2 \simeq  a_1(T/T_c-1)N^2+a_2 J N^4-BN+{\cal O}(N^6),
\end{equation}
where the coefficients $a_1$ and $a_2$, and the critical temperature $T_c$ can be numerically determined by solving the equation $\hat{L}\rho_n(r)=0$. The details will be presented in \cite{CaiYang2}.
One can see that the free energy density is quite similar to the one of the Ginzburg-Landau theory for the paramagnetism/ferromagnetism phase transition.~\footnote{According to the AdS/CFT dictionary, one should view $\rho_-$ as the magnetic moment in the dual field theory.  From the form (\ref{Gprho6}) of free energy, we think it is more natural to view $N$ rather than $\rho_-$ as the magnetic moment in this paper since it is $N$ which is coupled to the external field $B$ rather than $\rho_-$. In fact we can also express $N$ as a function of $\rho_-$.  In the symmetric phase with $B =0$, both $N$ and $\rho_-$ are zero, and in the condensed phase close to the critical temperature, $\rho_-$ and $N$ are indeed proportional to each other with the same critical behavior. Far from the critical temperature, the nonlinear relation between $N$ and $\rho_-$ is due to the self-interaction term $V(M^{\mu\nu})$ of the tenor field in (\ref{eomFM}).}

{\emph {The spontaneous magnetization and the response to the external magnetic field.}}--
In order to study the behavior of spontaneous magnetization and the response to the external magnetic field $B$, we have to solve the equations~\eqref{rhop} numerically. Since our focus is on the behavior of magnetic moment near the critical temperature, we pay attention on the equation of $\rho$. As an example, we take $J=-1,m^2=-3$ in equations~\eqref{rhop}. The other values for $J$ and $m^2$ satisfying the restriction~\eqref{m2} and $J<0$ give the qualitatively same results.

\begin{figure}[h!]
\includegraphics[width=0.3\textwidth]{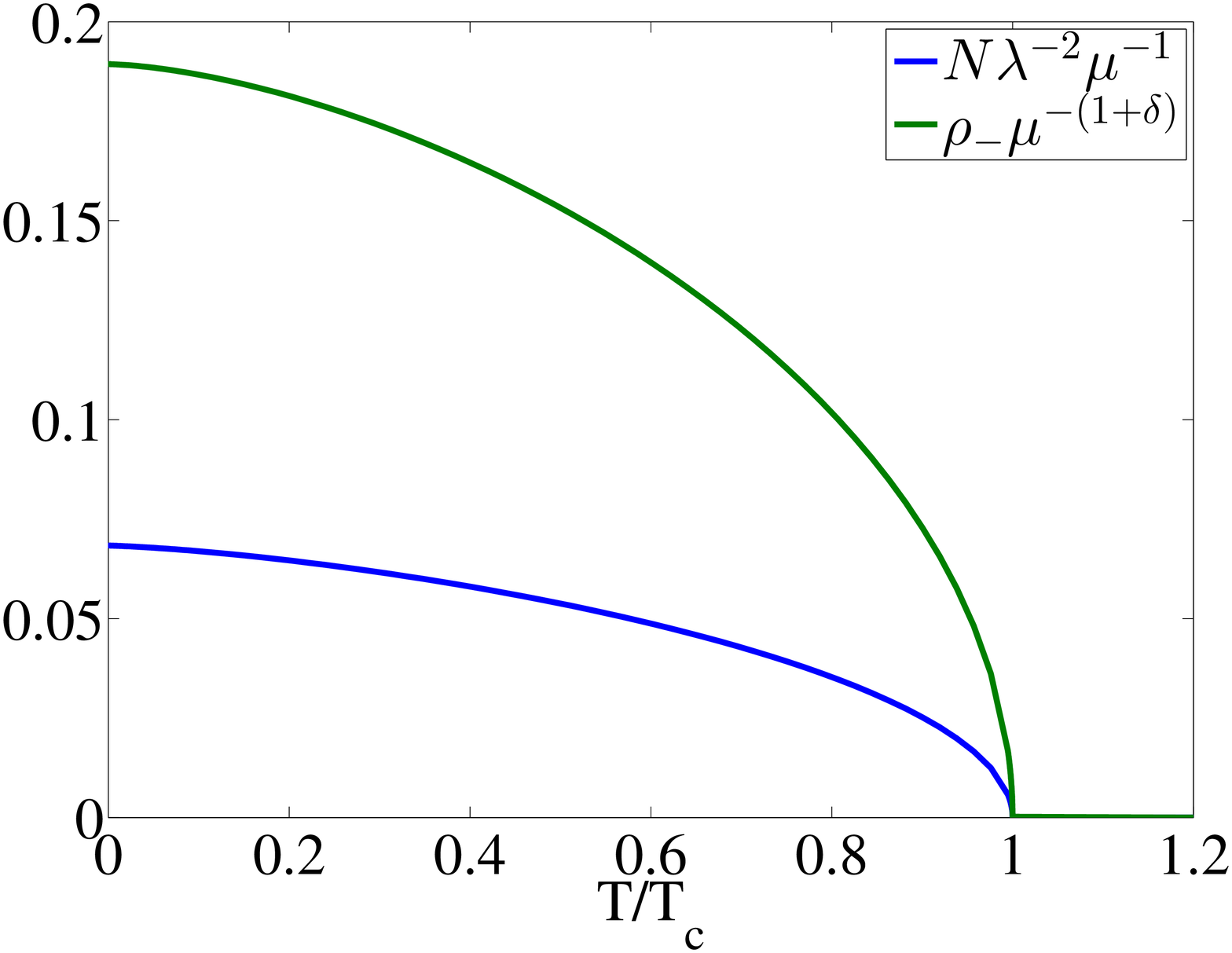}
\caption{The magnetic moment as a function of temperature. Here the critical temperature is $T_c/\mu\simeq0.00915$.}
\label{TM3}
\end{figure}

Firstly, we consider the spontaneous magnetization of the system. For this, we set $B=0$. Solving \eqref{rhop} with
requirement \eqref{delta1},
 we find there exists a critical temperature $T_c$. When $T<T_c$, the nontrivial solution $\rho\neq0$ appears. The magnetic moment $N$ as a function of temperature is plotted in Figure~\ref{TM3}. In this figure we also plot the behavior of $\rho_-$. Indeed we can see that near
 the critical point, $N$ and $\rho_-$ have the same behavior.

By fitting this curve near the critical temperature, we find that there is a square root behavior for the magnetic moment versus temperature, which is a typical behavior for a second order phase transition. Specifically, for $J=-1,m^2=-3$ we have
\begin{equation}\label{MT1}
N^2/\lambda^4\mu^2 \simeq 6.5054\times 10^{-3}  (1-T/T_c).
\end{equation}
This gives the critical exponent $1/2$, the same as the one from mean field theory. Note that if we take $\rho_-$ as the order parameter, it
has the same critical behavior near the critical temperature.  The second order phase transition can also be checked by computing the free energy of the system. It is easy to see that the free energy is continuous at the critical temperature. Thus we see that when $T<T_c$ the magnetic moment occurs spontaneously, which leads to the breaking of time reversal symmetry spontaneously. It is natural to expect that this system enters into a ferromagnetic phase as $T<T_c$.

Next let us turn on the external magnetic field $B$ to examine the response of magnetic moment $N$. This can be described by magnetic susceptibility density $\chi$, defined as
\begin{equation}\label{suscep}
\chi=\lim_{B\rightarrow0} \left. \frac{\partial N}{\partial B}\right |_{T}.
\end{equation}
In this model, the magnetic susceptibility comes from two parts, one is from the background black hole and the other from the polarization field. The pure dyonic Reissner-Nordstr\"om-AdS black hole corresponds to a diamagnetic material~\cite{Denef:2009yy}. Although in the probe limit, the contribution from the polarization field is suppressed by~$\lambda^2$,  near the critical temperature, we find that the part from the polarization field dominates for any nonzero $\lambda^2$ as this part is divergent at the critical temperature.
\begin{figure}
\includegraphics[width=0.22\textwidth]{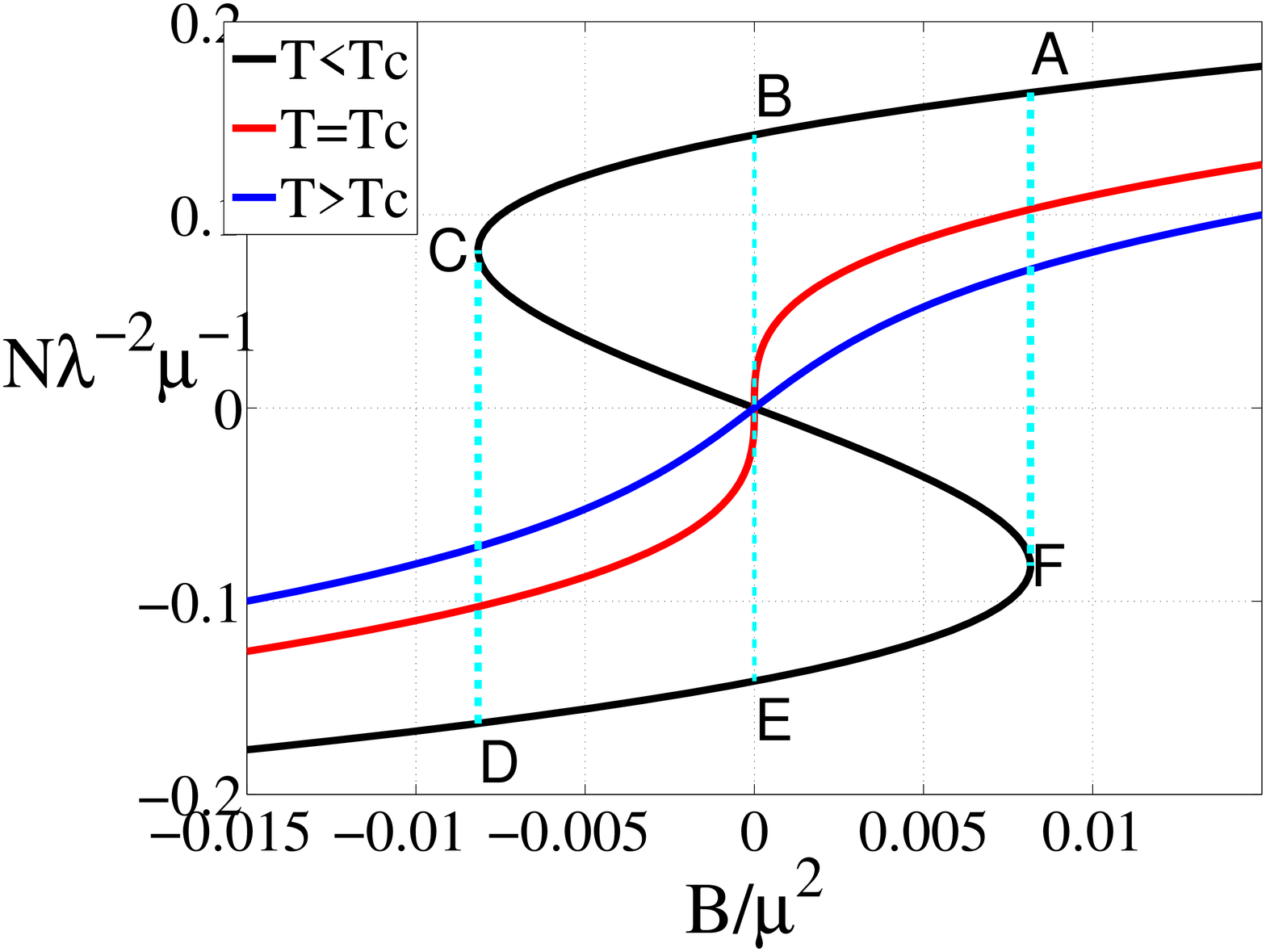}
\includegraphics[width=0.22\textwidth]{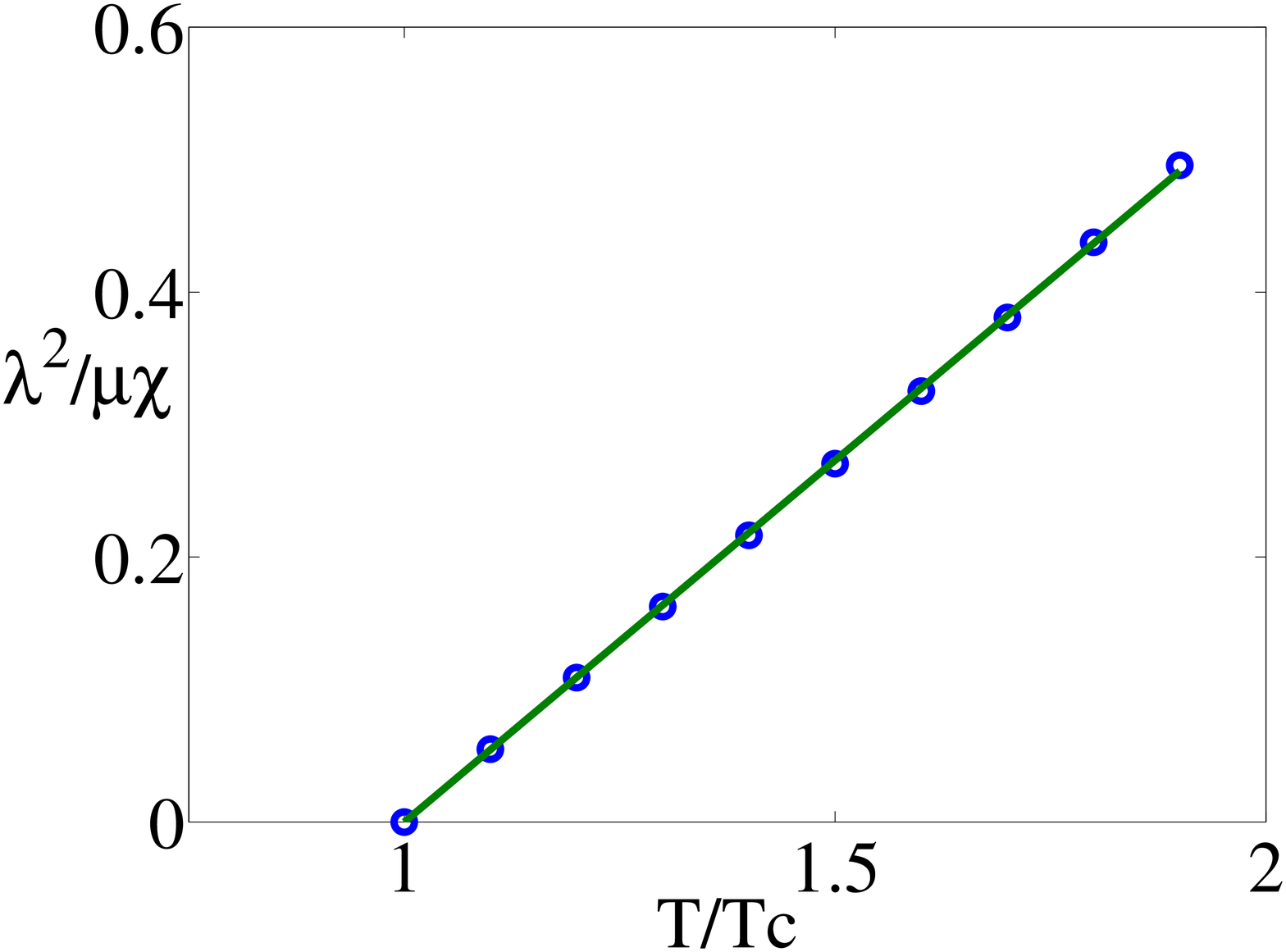}
\caption{Left: The magnetic moment with respect to external magnetic field $B$ in the different temperature. Right: The magnetic susceptibility as a function of temperature.}
\label{TM5}
\end{figure}
In the right plot of Figure~\ref{TM5}, we show the magnetic susceptibility as a function of temperature and find it satisfies the Curie-Weiss law of ferromagnetism in the region of $T>T_c$:
\begin{equation}\label{chi1}
\chi^{-1}\lambda^2/\mu \simeq  0.5463 (T/T_c-1).
\end{equation}
 Now we can conclude that the dual system is in a paramagnetic phase in high temperatures and a ferromagnetic phase in low temperatures. The model~\eqref{LM2} can describe a paramagnetism/ferromagnetism phase transition.

In the left plot of Figure~\ref{TM5}, we show the magnetic moment with respect to external magnetic field $B$ in the different temperatures. In the case of $T<T_c$, we can see that the magnetic moment is not single valued. The parts DE and BA are stable, which can be realized in the external field. The part CF is unstable which cannot exist in the realistic system. The parts EF and CB are metastable states, which may exist in some intermediate processes and can be observed in experiment. When the external field continuously changes between $-B_{max}$ and $B_{max}$ periodically, the metastable states of magnetic moment can appear. Thus we  see a hysteresis loop A-B-C-D-E-F-A. This corresponds to the hysteresis loop in the single magnetic domain.

{\emph {Summary and Discussion.}}--We have built a simple $(3+1)$-dimensional bulk gravity model which can reproduce some salient properties of a $(2+1)$-dimensional paramagnetism/ferromagnetism phase transition. In the case of zero external magnetic field, the model can break the time reversal symmetry spontaneously in low temperatures and realize the paramagnetism/ferromagnetism phase transition. The critical exponent agrees with the one from mean field theory. In the case of nonzero magnetic field, the model realizes the hysteresis loop of single magnetic domain and the magnetic susceptibility satisfies the Curie-Weiss law. Furthermore, let us mention here that the relation (\ref{MT1}) for the magnetic momentum  and (\ref{chi1}) for the magnetic susceptibility can also be obtained through the off shell free energy (\ref{Gprho6})~\cite{CaiYang2}.

Here some remarks are in order. In the ansatz (\ref{Mcomp}) we have not considered the components $M_{tx}$ and $M_{ty}$. In fact, these two components can be viewed as dual field of the order parameter in the paraelectric/ferroelectric phase transition in dielectric  materials. The spontaneous electric polarization can be realized in this model, which will be studied in a forthcoming paper~\cite{CaiYang2}.  As a result, our model provides a unified holographic description for the paramagnetism/ferromagnetism and paraelectric/ferroelectric phase transitions.
In addition, we can extend this model by including two different polarization tensors in order to form two different magnetic-ordered structures. By adding some suitable interaction between them, one can expect that the antiferromagnetism and ferrimagnetism phases will appear~\cite{CaiYang}. In particular, we would like to stress that this paper is just a first step to stress magnetic orders in some strongly correlated systems. By combining this model with some holographic superconductor models, one can explore the competition and coexistence between magnetic-ordered phase and superconducting phase in high temperature superconducting materials.

{\emph {Acknowledgements}}--The authors thank L Li and H Liu for helpful comments, discussions and suggestions. This work was supported in part by the National Natural Science Foundation of China (No.10821504, No.11035008, No.11375247, and No.11435006 ).



\end{document}